%Paper: hep-th/9406069
%From: cvj@guinness.ias.edu (Clifford Johnson)
%Date: Sun, 12 Jun 94 23:46:26 EDT
%Date (revised): Tue, 21 Jun 94 22:26:43 EDT
%Date (revised): Thu, 23 Jun 94 16:57:04 EDT

%%%%%%%%%%%%%%%%  preprint and letter macro  %%%%%%%%%%%%%%%%%

\hsize=6.0truein
\vsize=8.5truein
\voffset=0.25truein
\hoffset=0.1875truein%would be 0.25 except our laser printer is off by 1/16in
\tolerance=1000
\hyphenpenalty=500
\def\monthintext{\ifcase\month\or January\or February\or
   March\or April\or May\or June\or July\or August\or
   September\or October\or November\or December\fi}

%%%%%%%%%%%%%%%%%  Twelve point text font  %%%%%%%%%%%%%%%%%%%

\font\tenrm=cmr10 scaled \magstep1   \font\tenbf=cmbx10 scaled \magstep1
\font\sevenrm=cmr7 scaled \magstep1  
\font\fiverm=cmr5 scaled \magstep1   

\font\teni=cmmi10 scaled \magstep1   \font\tensy=cmsy10 scaled \magstep1
\font\seveni=cmmi7 scaled \magstep1  \font\sevensy=cmsy7 scaled \magstep1
\font\fivei=cmmi5 scaled \magstep1   \font\fivesy=cmsy5 scaled \magstep1

\font\tentt=cmtt10 scaled \magstep1
\font\tenit=cmti10 scaled \magstep1
\font\tensl=cmsl10 scaled \magstep1

\def\twelvepoint{\def\rm{\fam0\tenrm}
   \textfont0=\tenrm \scriptfont0=\sevenrm \scriptscriptfont0=\fiverm
   \textfont1=\teni  \scriptfont1=\seveni  \scriptscriptfont1=\fivei
   \textfont2=\tensy \scriptfont2=\sevensy \scriptscriptfont2=\fivesy
   \textfont\itfam=\tenit \def\it{\fam\itfam\tenit}
   \textfont\ttfam=\tentt \def\tt{\fam\ttfam\tentt}
   \textfont\bffam=\tenbf \def\bf{\fam\bffam\tenbf}
   \textfont\slfam=\tensl \def\sl{\fam\slfam\tensl} \rm
   %Essentially I changed all dimensions to 1.2 times as large as in plain tex
   \hfuzz=1pt\vfuzz=1pt%much more than plain tex's value
   \setbox\strutbox=\hbox{\vrule height 10.2pt depth 4.2pt width 0pt}
   \parindent=24pt\parskip=1.2pt plus 1.2pt
   \topskip=12pt\maxdepth=4.8pt\jot=3.6pt
   \normalbaselineskip=14.4pt\normallineskip=1.2pt
   \normallineskiplimit=0pt\normalbaselines
   \abovedisplayskip=13pt plus 3.6pt minus 5.8pt
   \belowdisplayskip=13pt plus 3.6pt minus 5.8pt
   \abovedisplayshortskip=-1.4pt plus 3.6pt
   \belowdisplayshortskip=13pt plus 3.6pt minus 3.6pt
   %plain tex's value for belowdisplayshortskip looked terrible
   \topskip=12pt \splittopskip=12pt
   \scriptspace=0.6pt\nulldelimiterspace=1.44pt\delimitershortfall=6pt
   \thinmuskip=3.6mu\medmuskip=3.6mu plus 1.2mu minus 1.2mu
   \thickmuskip=4mu plus 2mu minus 1mu%reduced these plain tex values
   \smallskipamount=3.6pt plus 1.2pt minus 1.2pt
   \medskipamount=7.2pt plus 2.4pt minus 2.4pt
   \bigskipamount=14.4pt plus 4.8pt minus 4.8pt}

\twelvepoint

%%%%%%%%%%%%%%%%% Definitions for Preprints %%%%%%%%%%%%%%%%%%

% title page title font

\font\titlerm=cmr10 scaled \magstep3
\font\titlerms=cmr10 scaled \magstep1 %\font\titlermss=cmr8
\font\titlei=cmmi10 scaled \magstep3  %math italic for title
\font\titleis=cmmi10 scaled \magstep1 %\font\titleiss=cmmi8
\font\titlesy=cmsy10 scaled \magstep3 	%math symbols for title
\font\titlesys=cmsy10 scaled \magstep1  %\font\titlesyss=cmsy8
\font\titleit=cmti10 scaled \magstep3	%text italic for title
\skewchar\titlei='177 \skewchar\titleis='177 %\skewchar\titleiss='177
\skewchar\titlesy='60 \skewchar\titlesys='60 %\skewchar\titlesyss='60

\def\titlefont{\def\rm{\fam0\titlerm}% switch to title font
   \textfont0=\titlerm \scriptfont0=\titlerms %\scriptscriptfont0=\titlermss
   \textfont1=\titlei  \scriptfont1=\titleis  %\scriptscriptfont1=\titleiss
   \textfont2=\titlesy \scriptfont2=\titlesys %\scriptscriptfont2=\titlesyss
   \textfont\itfam=\titleit \def\it{\fam\itfam\titleit} \rm}

% title page macros

\def\preprint#1{\baselineskip=19pt plus 0.2pt minus 0.2pt \pageno=0
   \begingroup%use with \draft or \date to end group
   \nopagenumbers\parindent=0pt\baselineskip=14.4pt\rightline{#1}}
\def\title#1{
   \vskip 0.9in plus 0.45in
   \centerline{\titlefont #1}}
\def\secondtitle#1{}%set up this some time
\def\author#1#2#3{\vskip 0.9in plus 0.45in
   \centerline{{\bf #1}\myfoot{#2}{#3}}\vskip 0.12in plus 0.02in}
\def\secondauthor#1#2#3{}%set up this some time
\def\addressline#1{\centerline{#1}}
\def\abstract{\vskip 0.7in plus 0.35in
	\centerline{\bf Abstract}
	\smallskip}
\def\finishtitlepage#1{\vskip 0.8in plus 0.4in
   \leftline{#1}\supereject\endgroup}

\def\date#1{\finishtitlepage{#1}}

\def\nolabels{\def\eqnlabel##1{}\def\eqlabel##1{}\def\figlabel##1{}%
	\def\reflabel##1{}}
\def\writelabels{\def\eqnlabel##1{%
	{\escapechar=` \hfill\rlap{\hskip.11in\string##1}}}%
	\def\eqlabel##1{{\escapechar=` \rlap{\hskip.11in\string##1}}}%
	\def\figlabel##1{\noexpand\llap{\string\string\string##1\hskip.66in}}%
	\def\reflabel##1{\noexpand\llap{\string\string\string##1\hskip.37in}}}
\nolabels

%  tagged section numbers

\global\newcount\secno \global\secno=0
\global\newcount\meqno \global\meqno=1
\global\newcount\subsecno \global\subsecno=0

\font\secfont=cmbx12 scaled\magstep1

\def\section#1{\global\advance\secno by1
   \xdef\secsym{\the\secno.}
   \global\subsecno=0
   \global\meqno=1\bigbreak\medskip
   \noindent{\secfont\the\secno. #1}\par\nobreak\smallskip\nobreak\noindent}
%\xdef\secsym{}

\def\subsection#1{\global\advance\subsecno by1
    %\xdef\secsym{\the\subsecno}
\medskip
\noindent
{\bf\the\secno.\the\subsecno\ #1}
\par\medskip\nobreak\noindent}
%\xdef\secsym{}

\def\newsec#1{\global\advance\secno by1
   \xdef\secsym{\the\secno.}
   \global\meqno=1\bigbreak\medskip
   \noindent{\bf\the\secno. #1}\par\nobreak\smallskip\nobreak\noindent}
\xdef\secsym{}

\def\appendix#1#2{\global\meqno=1\xdef\secsym{\hbox{#1.}}\bigbreak\medskip
\noindent{\bf Appendix #1. #2}\par\nobreak\smallskip\nobreak\noindent}

%         equations

\def\eqnn#1{\xdef #1{(\secsym\the\meqno)}%
	\global\advance\meqno by1\eqnlabel#1}
\def\eqna#1{\xdef #1##1{\hbox{$(\secsym\the\meqno##1)$}}%
	\global\advance\meqno by1\eqnlabel{#1$\{\}$}}
\def\eqn#1#2{\xdef #1{(\secsym\the\meqno)}\global\advance\meqno by1%
	$$#2\eqno#1\eqlabel#1$$}

%			 footnotes

\def\myfoot#1#2{{\baselineskip=14.4pt plus 0.3pt\footnote{#1}{#2}}}
%sequentially numbered footnotes
\global\newcount\ftno \global\ftno=1
\def\foot#1{{\baselineskip=14.4pt plus 0.3pt\footnote{$^{\the\ftno}$}{#1}}%
	\global\advance\ftno by1}

%         references

\global\newcount\refno \global\refno=1
\newwrite\rfile

\def\ref{[\the\refno]\nref}
\def\nref#1{\xdef#1{[\the\refno]}\ifnum\refno=1\immediate
	\openout\rfile=refs.tmp\fi\global\advance\refno by1\chardef\wfile=\rfile
	\immediate\write\rfile{\noexpand\item{#1\ }\reflabel{#1}\pctsign}\findarg}
%	horrible hack to sidestep tex \write limitation
\def\findarg#1#{\begingroup\obeylines\newlinechar=`\^^M\passarg}
	{\obeylines\gdef\passarg#1{\writeline\relax #1^^M\hbox{}^^M}%
	\gdef\writeline#1^^M{\expandafter\toks0\expandafter{\striprelax #1}%
	\edef\next{\the\toks0}\ifx\next\null\let\next=\endgroup\else\ifx\next\empty%

\else\immediate\write\wfile{\the\toks0}\fi\let\next=\writeline\fi\next\relax}}
	{\catcode`\%=12\xdef\pctsign{%}}
\def\striprelax#1{}

\def\semi{;\hfil\break}
\def\addref#1{\immediate\write\rfile{\noexpand\item{}#1}} %now unnecessary

\def\listrefs{\vfill\eject\immediate\closeout\rfile
   {{\secfont References}}\bigskip{\frenchspacing%
   \catcode`\@=11\escapechar=` %
   \input refs.tmp\vfill\eject}\nonfrenchspacing}

\def\startrefs#1{\immediate\openout\rfile=refs.tmp\refno=#1}

%		and finally, figures:

\global\newcount\figno \global\figno=1
\newwrite\ffile
\def\fig{\the\figno\nfig}
\def\nfig#1{\xdef#1{\the\figno}\ifnum\figno=1\immediate
	\openout\ffile=figs.tmp\fi\global\advance\figno by1\chardef\wfile=\ffile
	\immediate\write\ffile{\medskip\noexpand\item{Fig.\ #1:\ }%
	\figlabel{#1}\pctsign}\findarg}

\def\listfigs{\vfill\eject\immediate\closeout\ffile{\parindent48pt
	\baselineskip16.8pt{{\secfont Figure Captions}}\medskip
	\escapechar=` \input figs.tmp\vfill\eject}}

%%%%%%%%%%%%%%%%%%%%%%%%%%%%%%%%%%%%%%%%%%%%%%%%%%%%%%%%%%%%%%%%%%%%%%%%%%%%%%%
\def\noblackbox{\overfullrule=0pt}
\def\inv{^{\raise.18ex\hbox{${\scriptscriptstyle -}$}\kern-.06em 1}}
\def\dup{^{\vphantom{1}}}
\def\Dsl{\,\raise.18ex\hbox{/}\mkern-16.2mu D} %this one can be subscripted
\def\dsl{\raise.18ex\hbox{/}\kern-.68em\partial}
\def\slash#1{\raise.18ex\hbox{/}\kern-.68em #1}
\def\lspace{}
\def\lbspace{}
\def\boxeqn#1{\vcenter{\vbox{\hrule\hbox{\vrule\kern3.6pt\vbox{\kern3.6pt
	\hbox{${\displaystyle #1}$}\kern3.6pt}\kern3.6pt\vrule}\hrule}}}
\def\mbox#1#2{\vcenter{\hrule \hbox{\vrule height#2.4in
	\kern#1.2in \vrule} \hrule}}  %e.g. \mbox{.1}{.1}
%matters of taste
%\def\tilde{\widetilde}
\def\bar{\overline}
\def\e#1{{\rm e}^{\textstyle#1}}
\def\del{\partial}
\def\curly#1{{\hbox{{$\cal #1$}}}}
\def\curlyD{\hbox{{$\cal D$}}}
\def\curlyL{\hbox{{$\cal L$}}}
\def\vev#1{\langle #1 \rangle}
\def\psibar{\overline\psi}
\def\lform{\hbox{$\sqcup$}\llap{\hbox{$\sqcap$}}}
\def\darr#1{\raise1.8ex\hbox{$\leftrightarrow$}\mkern-19.8mu #1}
\def\half{{\textstyle{1\over2}}} %puts a small half in a displayed eqn
\def\roughly#1{\ \lower1.5ex\hbox{$\sim$}\mkern-22.8mu #1\,}
\def\MSbar{$\bar{{\rm MS}}$}
%%%%%%%%%%%%%%%%%%%%%%%%%%%%%%%%%%%%%%%%%%%%%%%%%%%%%%%%%%%%%%
\hyphenation{di-men-sion di-men-sion-al di-men-sion-al-ly}

\parindent=0pt
\parskip=5pt

\def\Tr{{\rm Tr}}
\def\det{{\rm det}}
\def\jump{\hskip1.0cm}
\def\wzw{Wess--Zumino--Witten}
\def\Az{A_z}
\def\Azb{A_{\bar{z}}}
\def\lr{\lambda_R}
\def\ll{\lambda_L}
\def\lrb{\bar{\lambda}_R}
\def\llb{\bar{\lambda}_L}
\font\top = cmbxti10 scaled \magstep1

\def\d{\partial_z}
\def\db{\partial_{\bar{z}}}
\def\rline{{{\rm I}\!{\rm R}}}
\def\tl{t_L}
\def\tr{t_R}
\def\ie{{\it i.e.,}\ }
\def\cam{{\cal M}}
\def\cak{{\cal K}}

\preprint{
\vbox{\rightline{IASSNS--HEP--94/50}
\vskip2pt\rightline{McGill/94-28}
\vskip2pt\rightline{hep-th/9406069}
\vskip2pt\rightline{12th June 1994.}
}
}
\vskip -2.0cm
\title{Taub--NUT Dyons in Heterotic String Theory.}
\vskip 0.5cm
\vskip -1.5cm
\author{\bf Clifford V. Johnson\myfoot{$^*$}{\rm e-mail: cvj@guinness.ias.edu}
and Robert C. Myers\myfoot{$^\dagger$}{\rm e-mail: %
rcm@hep.physics.mcgill.ca}}{}{}
\vskip 0.7cm
\addressline{\it $^*$School of Natural Sciences}
\addressline{\it Institute for Advanced Study}
\addressline{\it Olden Lane}
\addressline{\it Princeton, NJ 08540 U.S.A.}
\bigskip
\addressline{\it $^{\dagger}$Department of Physics, McGill University}
\addressline{\it Ernest Rutherford Physics Building}
\addressline{\it 3600 University Street}
\addressline{\it Montr\'eal, Qu\'ebec, Canada H3A 2T8.}
\vskip -1.0cm
\abstract

Starting with the Taub--NUT solution to Einstein's equations, together with a
constant dilaton, a  dyonic Taub--NUT solution of low energy heterotic
string theory  with non--trivial dilaton, axion and $U(1)$ gauge fields is
constructed by employing $O(1,1)$ transformations. The
electromagnetic dual of this  solution is constructed, using $SL(2,\rline)$
transformations. By an appropriate change to scaled variables, the
extremal limit of the dual solution is shown to correspond to the low
energy limit of an exact conformal field theory presented previously.

\date{Revised Version, 21st June 1994.}
%\draft{}

\newsec{Introduction.}

Since it is hoped that string theory can provide a consistent theory of
quantum
gravity, there has been a great deal of interest in studying nontrivial
gravitational solutions of the string equations of
motion\ref\dark{G.T. Horowitz, ``The Dark Side of String Theory: Black
Holes and Black Strings,'' lectures at Trieste Spring School (1992)
hep-th/9210119.}. A great
deal of activity was stimulated by Witten's discovery\ref\witten{E. Witten,
{\sl Phys. Rev.} {\bf D44} (1991) 314.} that an $SL(2,\rline)/U(1)$ conformal
field theory describes a two-dimensional black hole.
Unfortunately, no such exact conformal field theory solutions
providing a complete description of a black hole in four dimensions
(including the asymptotically
flat regions) has been constructed. In a number of cases though,
the exact solutions describing the throat geometry of a four-dimensional
black hole carrying an extremal charge have been
found\ref\throat{S.B. Giddings, J. Polchinski and A. Strominger,
{\sl Phys. Rev.} {\bf D48} (1993) 5748; W. Nelson, ``Kaluza-Klein
Black Holes in String Theory,'' preprint hep-th/9312058; D.A. Lowe
and A. Strominger, ``Exact Four-Dimensional Dyonic Black Holes
and Bertotti-Robinson Spacetimes in String Theory,'' preprint
hep-th/9403186.}\ref\cvj{C.V. Johnson, ``Exact Models of Extremal Dyonic
4D Black
Hole Solutions of Heterotic String Theory,'' hep-th/9403192, to appear in
{\sl Phys. Rev.} {\bf D}}.
Amongst the exact solutions constructed in ref.\cvj,
the low energy limit of one solution had a time coordinate which was compact,
and which formed a nontrivial fibre bundle over the
spatial two-sphere. In that paper, it was conjectured that
this  conformal field theory should correspond to the throat limit
of an extremal Taub--NUT dyon in heterotic string
theory.

In the present paper, we confirm this conjecture by constructing
the Taub--NUT dyons, which are solutions of the
low-energy effective string equations. Our construction
relies on the two remarkable ``duality'' symmetries of the
effective string action, which provide a technique to
generate new solutions. In particular, for heterotic strings in
$d$-dimensions coupled to a gauge group of rank $p$, there
is a possible symmetry group of $O(d,d+p)$. These transformations are related
to the heralded string duality which relates large and small radius
compactifications. In section 2, we employ these transformations to produce
a Taub--NUT solution carrying nontrivial electric and magnetic
charges from the vacuum Taub--NUT solution.
In four dimensions, there is also an $SL(2,\rline)$ symmetry of the low--energy
equations of motion. These transformations yield the strong--to--weak
coupling duality of four--dimensional string theory.
The electric and magnetic couplings of particles and backgrounds
are also exchanged under these transformations. In section 3,
this electromagnetic duality transformation is applied to the Taub--NUT
dyon.
In section 4, the extremal limit of these dyons is considered,
and we explicitly show that the low energy background fields of the exact
conformal field theory considered in ref.\cvj\ are recovered.
Section 5 provides a brief discussion of our results.

First let us establish our conventions for the normalization of the
fields. We write the four-dimensional low energy effective action
for the heterotic string as
\eqn\actg{I=
\int d^4x\,\sqrt{-G}e^{-\Phi}\,\left(R(G)+(\nabla\Phi)^2-{1\over12}H^2
-{1\over8}F^2+\ldots\right)\ \ .}
Here, the three form $H$ has components
\eqn\hdef{H_{\mu\nu\rho}=\partial_\mu B_{\nu\rho}+\partial_\nu
B_{\rho\mu}+\partial_\rho B_{\mu\nu}-\omega(A)_{\mu\nu\rho},}
where
\eqn\chernsimons{\omega(A)_{\mu\nu\rho}={1\over4}(A_\mu F_{\nu\rho}+A_\nu
F_{\rho\mu}+A_\rho F_{\mu\nu})}
is the Chern--Simons three form for the U(1) gauge
field\foot{We have dropped the contribution to $H$ from the Lorentz
Chern--Simons three form, as being higher order in the $\alpha'$
expansion. Note also that we have rescaled the gauge fields from
their standard normalization by a factor of $1/\sqrt{\alpha'}$.
Therefore thinking of our background solutions  as valid for
small $\alpha'$ is equivalent to thinking of them as carrying
large (electric and magnetic) charges.}. The $U(1)$ field
strength is $F_{\mu\nu}=\partial_\mu A_\nu-\partial_\nu A_\mu$.
In eq.\actg, the action
is written in terms of the string or sigma model metric $G_{\mu\nu}$.
It is also convenient to introduce the Einstein metric,
$g_{\mu\nu}=\exp(-{\Phi})G_{\mu\nu}$, for which the effective action
becomes
\eqn\actgg{I=
\int d^4x\,\sqrt{-g}\,\left(R(g)
-{1\over2}(\nabla\Phi)^2-{1\over12}e^{-2\Phi}H^2
-{1\over8}e^{-\Phi}F^2+\ldots\right)\ \ .}
Our notation will always be to write the Einstein metric with
a lower case $g$, and the string metric with an upper case $G$.
In displaying the line elements, $ds^2_E$ and $ds^2_S$ will
denote that for the Einstein and string metrics, respectively.
(No subscript will be used if the two metrics are the same.)

\newsec{The dyonic solution via $O(1,1)$}

The Taub--NUT solution\ref\TaubNUT{A.H. Taub, {\sl Ann. Math.} {\bf 53}
(1951) 472; E. Newman, L. Tamborino and T. Unti, {\sl J. Math. Phys.}
{\bf 4} (1963) 915.}:
\eqn\taubnut{ds^2 =-f_1
(dt+2l\cos\!\theta\, d\phi)^2
+f_1^{-1}dr^2+(r^2+l^2)(d\theta^2+
\sin^2\!\theta\, d\phi^2)}
where \eqn\fone{f_1=1-2{Mr+l^2\over r^2+l^2},} is a solution to the
vacuum Einstein's equations $R_{\mu\nu}=0.$ This metric
\taubnut\ is invariant under four Killing
symmetries\ref\misner{C.W. Misner, {\sl J. Math. Phys.} {\bf 4} (1963) 924;
C.W. Misner and A.H. Taub, {\sl Sov. Phys. JETP} {\bf 28} (1969) 122;
C.W. Misner in {\sl Relativity Theory and Astrophysics I: Relativity
and Cosmology}, ed. J. Ehlers, Lectures in Applied Mathematics {\bf 8}
(American Mathematical Society, 1967) 160.}.
The first of these is simply time translations. The remaining
three act as $SO(3)$ rotations on the angular coordinates, but also
involve time translations to preserve the differential
$dt+2l\cos\theta\,d\phi$. Thus the orbit of a point under these
``rotations'' is in fact three-dimensional (when $l\ne0$).
Consistency requires $t$ to have  periodicity
  $8l\pi$ as can be seen by taking care to ensure that there are no
conical curvature singularities at $\theta=0$ or $\pi$.
Surfaces of constant radius then have the topology of a three-sphere,
in which there is a Hopf fibration of the $S^1$ of time over the spatial
$S^2$. The metric \taubnut\ is also singular at
$r=r_\pm=M\pm(M^2+l^2)^{1/2}$, but a nonsingular extension across these
null surfaces can be found just as at the event horizon of a
black hole\misner. Unfortunately, the periodicity of the time
coordinate prevents an interpretation of the Taub--NUT solution
as a black hole. Of course if $l=0$, one recovers the usual
Schwarzschild black hole geometry, and $t$ is not periodically
identified. The interior region (\ie $r_-<r<r_+$) may be
interpreted as a cosmological solution for a universe with the spatial
topology $S^3$.

Combined with vanishing gauge and antisymmetric tensor fields,
as well as a constant dilaton --- we choose $\Phi=0$ for simplicity ---
the Taub--NUT metric \taubnut\ is a solution of the low energy string
equations to leading order in the $\alpha^\prime$ expansion.
We wish to generalize this solution by introducing a nontrivial
electromagnetic field. A charged Taub--NUT solution of the
Einstein-Maxwell equations is known\ref\brill{D. Brill, {\sl Phys.
Rev.} {\bf 133} (1964) B845.}, but it will not provide a solution
of the string equations because of the nontrivial dilaton and axion
couplings to the gauge field
in the effective action \actg. These couplings though
lead to a remarkable symmetry of the action, which allows us to
produce a new dyonic Taub--NUT solution.

In refs.\ref\duals{G. Veneziano, {\sl Phys. Lett.} {\bf B265}
(1991) 287; K. Meissner and G. Veneziano, {\sl Phys. Lett.} {\bf B267}
(1991) 33; A. Sen, {\sl Phys. Lett.} {\bf B274} (1992) 34;
M. Gasperini, J. Maharana and G. Veneziano, {\sl Phys. Lett.}
{\bf B272} (1991) 277.}\ref\puregauge{A. Sen, {\sl Phys. Lett.} {\bf B271}
(1991) 295.}, it was shown that there exists
an $O(d-1,1)\otimes O(d-1,1)$ symmetry of the space of solutions of the low
energy field equations when the solutions are independent of $d$ of the
spacetime coordinates. (Here, the time coordinate has been included, with a
Minkowskian signature metric.) In ref.\ref\senone{S.F. Hassan and A. Sen,
{\sl Nucl. Phys.} {\bf B375} (1992) 103.}, the
extension of these results to the case of heterotic string theory was
presented. For solutions which are independent of $d$ of the spacetime
coordinates and for which the background gauge field lies in a subgroup
commuting with $p$ of the $U(1)$ generators of the gauge group, there is an
$O(d-1,1)\otimes O(d+p-1,1)$ symmetry of the field equations.
In fact, there is a larger symmetry group
$O(d,d+p)$, however those transformations which are not in the
$O(d-1,1)\otimes O(d+p-1,1)$ subgroup correspond to pure
gauge transformations\puregauge. The remaining transformations
provide a powerful technique to generate new solutions, and by
which  to explore the space of solutions.

Regarding the solution \taubnut\ with $\Phi=0$ as an uncharged
solution to heterotic string theory, we supplement
the four spacetime coordinates with one extra ``chiral'' coordinate,
$X$, which allows a $U(1)$ gauge field background to be
considered.
Now the starting solution is independent of the $t$ and $\phi$ coordinates,
as well as having no gauge fields, which therefore trivially commutes with the
$U(1)$ carried by  the
$X$ direction. In general,
there is the possibility of doing $O(1,1)\otimes O(2,1)$
transformations to generate more solutions.
For the concerns here it is not of interest to include the possibility of
boosts in the $\phi$ direction, as this will break the ``rotation''
symmetries of the Taub--NUT solution discussed above.

So an $O(1,1)$ boost in the gauge--time directions will be performed. Using the
notation of ref.\senone, the metric \taubnut\ is used to form the $9\times
9$ matrix $\cam$
\eqn\sendef{
\cam=\pmatrix{\cak_-^TG^{-1}\cak_-&\cak_-^TG^{-1}\cak_+&-\cak_-^TG^{-1}A\cr
              \cak_+^TG^{-1}\cak_-&\cak_+^TG^{-1}\cak_+&-\cak_+^TG^{-1}A\cr
              -A^TG^{-1}\cak_-&-A^TG^{-1}\cak_+&A^TG^{-1}A\cr}  }
where
\eqn\Kmatrix{\eqalign{
(\cak_\pm)_{\mu\nu}&=-B_{\mu\nu}-G_{\mu\nu}-{1\over4}A_\mu A_\nu
\pm\eta_{\mu\nu}\cr
\eta_{\mu\nu}&={\rm diag}(1,1,1,-1)\cr}}
and $T$ denotes matrix transpose.
The boost matrix for the gauge--time directions
\eqn\Mmatrix{\Omega=\pmatrix{I_7&0&0\cr0&x&
\sqrt{x^2-1}\cr0&\sqrt{x^2-1}&x}} is defined, where
$I_7$ is a $7\times7$ identity matrix, and $x^2\ge 1.$
The equations of motion are invariant under
\eqn\newm{\cam\to \cam'=\Omega \cam\Omega^T.}
{}From the matrix $\cam'$, the new metric $G^\prime$, antisymmetric tensor
and gauge  fields may be extracted by following the definitions in \senone.
The new dilaton field is given by
\eqn\newdilaton{\Phi^\prime=-{1\over2}\log\left({{\rm det}G\over{\rm
det}G^\prime}\right)} where $G$ refers to the old string metric.
This gives a dyonic Taub--NUT solution to heterotic string theory:
\eqn\newmetrics{\eqalign{
ds^2_S =-{f_1\over f_2^2}(dt+(x+1)l\cos\!\theta\, d\phi)^2
+{f_1^{-1}dr^2}+(r^2+l^2)(d\theta^2+
\sin^2\!\theta\, d\phi^2),
}}
where
\eqn\where{\eqalign{f_1&=1-2{Mr+l^2\over r^2+l^2}\cr
{\rm and}\hskip 0.5cm f_2&=1+(x-1){Mr+l^2\over r^2+l^2}}}
together with:
\eqn\newfields{\eqalign{B_{t\phi}&={f_1\over f_2}
(x-1)\,l\cos\theta\cr
A_\phi&=-2{f_1\over f_2}
\sqrt{x^2-1}\,l\cos\theta\cr
A_t&=\sqrt{x^2-1}\left({1-f_1\over f_2}\right)\cr
e^{-\Phi}&=f_2.}}
The line element for the Einstein metric will be given by
$ds^2_E=f_2ds^2_S$. Applying the previous analysis of
the Taub--NUT solution to the present metric, one is again lead to
consider constant $r$ surfaces with topology $S^3$ in which
$t$ is identified with a period $4(x+1)l\pi$.
With $l=0$, one recovers an electrically
charged black hole solution discussed in ref.\ref\charged{G.W.
Gibbons, {\sl Nucl. Phys.} {\bf B207} (1982) 337;
G.W. Gibbons and K. Maeda, {\sl Nucl. Phys.} {\bf B298} (1988) 741;
D. Garfinkle, G. Horowitz and A. Strominger, {\sl Phys. Rev.}
{\bf D43} (1991) 3140; {\bf D45} (1992) 3888(E).}.

Examining the asymptotic behavior of $g_{tt}$ in the Einstein metric,
the mass $\mu$ of the solution is easily evaluated as
\eqn\mass{\mu={x+1\over2}M.}
The  non--zero components of the gauge field strength are $F_{rt},F_{r\phi}$
and  $F_{\theta\phi}$. One may be tempted to define
the magnetic and electric charges by integrating $F$ and its dual
over the spatial two--sphere at infinity. However because of the
topology of the constant $r$ surfaces, there are
in fact no nontrivial two spheres
on which to integrate. Hence we can only define the magnetic and electric
charges in terms of the asymptotic behavior of the electromagnetic fields
by analogy to that in an asymptotically flat space-time
(\ie $F_{tr}\simeq Q_E/r^2$ and $F_{\theta\phi}\simeq
Q_M\sin\theta$):\foot{Alternatively, the same charges would be determined
by considering the motion of point-like electric or magnetic
charges in the asymptotic region.}
\eqn\charge{\eqalign{Q_E
=2M\sqrt{x^2-1}\qquad{\rm and}\qquad
Q_M=2l\sqrt{x^2-1}.}}
Note that the physical
mass, and magnetic and electric charges of the solution are
completely independent, as they are independent
functions of the three
parameters $M$, $x$ and $l$.
By examining the asymptotic behavior of the dilaton and using
the analogy with an asymptotically flat space-time,
one may also define a non--vanishing dilaton charge from $e^\Phi
=1+{D\over r}$, which yields $D=-(x-1)M$.

\newsec{The electromagnetic  dual solution via $SL(2,\rline)$}

In four dimensions, the string equations of motion possess
another noteworthy symmetry
which may be employed as a further solution  generating technique.
This is a stringy electromagnetic duality
invariance\ref\frank{A. Shapere, S. Trivedi and F. Wilczek, {\it Mod.
Phys. Lett.} {\bf A6} (1991) 2677; A. Sen, {\it Nucl. Phys.}
{\bf B404} (1993) 109.}. This symmetry is also known as
$S$ duality since it relates solutions of strong and weak coupling.

To apply these transformations, one works in terms of the Einstein metric.
Then the scalar axion field,
$a$, must be determined from
\eqn\axiondef{
H_{\mu\nu\rho}=-e^{2\Phi}\epsilon_{\mu\nu\rho\kappa}\nabla^\kappa a}
where $\epsilon_{\mu\nu\rho\kappa}$ is the completely antisymmetric
tensor in four dimensions with $\epsilon_{tr\theta\phi}=\sqrt{-g}$.
Next define the complex scalar
\eqn\moredefone{\lambda=a+i\e{-\Phi}}
and the complex gauge field strengths
\def\tF{{\tilde F}}
\eqn\moredeftwo{(F_\pm)_{\mu\nu}=F_{\mu\nu}\pm {i\over2}
\epsilon_{\mu\nu\rho\kappa}F^{\rho\kappa}\ \ .}

The equations of motion are invariant under the transformations\frank
\eqn\duality{\eqalign{
T:\ (\lambda,F_+,F_-)&\to(\lambda+c,F_+,F_-)\cr
S:\ (\lambda,F_+,F_-)&\to(-{1\over\lambda},-\lambda F_+,-{\bar\lambda}F_-)\cr}}
where the Einstein metric is left invariant under both transformations.
In the above $\bar\lambda$ is the complex conjugate of $\lambda$, and $c$ in
a real constant\foot{Instanton
corrections\frank\ fix $c$ to be 1, and hence break the group to $SL(2,{\rm
Z})$.}.
Combined, these symmetry transformations generate the group
$SL(2,\rline)$.

For the solution presented in the previous section, the possible non--zero
components of $H$ are $H_{r\phi t}$ and $H_{\theta\phi t}$. After some algebra
the first is seen to vanish, yielding for the
the axion (discarding an integration constant):
\eqn\axion{a=(x-1)\,l{r-M\over r^2+l^2}.}
 Note that the fact that the axion has only  radial
dependence is consistent
with the full solution retaining the time translation
and rotation invariances of the original Taub--NUT solution.

After applying the $S$ duality transformation,
the new axion and dilaton fields are
\eqn\newaxidil{\e{-{\hat\Phi}}={f_2\over a^2+f_2^2},\,\,\quad
{\hat a}=-{a\over a^2+f_2^2}.} Combining the new dilaton with the (invariant)
Einstein metric gives the new string metric:
\eqn\newsigma{d{\hat s}^2_S=-{f_1\over f_2^2}(a^2+f_2^2)d\xi^2
+{(a^2+f_2^2)\over f_1}dr^2
+(a^2+f_2^2)(r^2+l^2)(d\theta^2+\sin^2\theta\,d\phi^2),}
where $d\xi=dt+(x+1)l\cos\theta\,d\phi$, and $f_1$ and $f_2$ are
given in eq.~\where.

\def\hF{{\hat F}}
The new gauge fields strengths are given by
\eqn\useful{
\hF_{\mu\nu}={1\over2}e^{-\Phi}\epsilon_{\mu\nu\rho\kappa}F^{\rho\kappa}-
a F_{\mu\nu}}
and one finds that the only only
 non--vanishing components of the $U(1)$ field strength are
$\hF_{rt},\hF_{r\phi}$ and  $\hF_{\theta\phi}$ from which  gauge potentials may
be derived:
\eqn\morepot{\eqalign{\hat{A}_\phi&=2\sqrt{x^2-1}\cos\theta \left({Mr^2+
((x-1)M^2+(x+1)l^2)r-Ml^2\over r^2+(x-1)Mr+xl^2}\right)\cr
\hat{A}_t&={2l\sqrt{x^2-1}(r-M)\over r^2+Mr(x-1)+xl^2}.}}
Together with the new axion and dilaton:
\eqn\newaxidil{\eqalign{{\hat a}&=-{(x-1)l(r-M)\over(r+M(x-1))^2+x^2l^2}\cr
\e{-{\hat\Phi}}&={r^2+Mr(x-1)+xl^2\over(r+M(x-1))^2+x^2l^2}}}
the new antisymmetric tensor field may be deduced as:
\eqn\newB{{\hat B}_{\phi t}=l(x-1)\cos\theta \left({ (r-M)^2+x(M^2+l^2)\over
r^2+Mr(x-1)+xl^2}\right).}

It is
straightforward to calculate the electric and magnetic charges as before:
\eqn\newcharges{\eqalign{{\hat Q}_E&=2l\sqrt{x^2-1}\cr
{\hat Q}_M&=-2M\sqrt{x^2-1}.}} Comparing this to the charges \charge\ of the
original dyon solution, it is noteworthy  that the
roles of $M$ and $l$ are simply
exchanged under duality.
The mass of the new solution is still \eqn\newmass{{\hat\mu}=\mu={x+1\over2}M}
 as the Einstein metric
remains invariant.
In this case, $l=0$ yields the magnetically charged black hole
solution of ref.\charged.

\newsec{The extremal limit and  an exact solution.}

Now for both solutions the charge to mass ratio is given by
\eqn\cmratio{{Q_E^2+Q_M^2\over\mu^2}=8\left({M^2+l^2\over M^2}\right){x-1\over
x+1}.}
In the pure black hole case $l=0$, the extremal limit is obtained when this
ratio is maximized, that is when $Q^2=8\mu^2$, where $Q=Q_E$ or $Q_M$ depending
upon whether the ($l=0$) solution is electric (section 2) or magnetic (section
3). (Note that our conventions differ from those of ref.\charged.)
This limit is achieved by taking the limit $M\to0$ and $x\to\infty$ while
holding the quantity  $m=xM$ a constant.
For non--zero $l$, it is clear that the choice $l\to0$  must be made while
holding the quantity $\lambda=xl$ finite. One way of seeing this is to simply
require a sensible limit for the line element
$d\xi=dt+(1+x)l\cos\theta\,d\phi$.

One of the interesting properties of charged string black hole metrics is that
at extremality the spatial geometry approaches that of an infinite throat of
constant
radius for increasing distance from the black hole.  This is because at
extremality the string metric
component $G_{rr}$
develops a double pole and therefore a measure of the proper distance $D$ to
travel  to the horizon at $r_H$  is:
\eqn\travel{D=\int_{r_0}^{r_1}{dr\over r-r_H}=
\log\left({r_1-r_H\over r_0-r_H}\right)}
which diverges as $r_0$ approaches $r_H$.
So the asymptotic throat geometry for constant time slices is
a manifestation of the fact that  the horizon is infinitely far away from
any point at finite $r$ in the outside region.

A careful way of approaching the extremal limit is to is to change variables so
as to send the asymptotically flat region away to infinity and define scaled
variables
which capture the physics near the horizon. To that end it is useful to set
$r=f(\sigma)$ where $f(\sigma)$ is determined only by the fact that a constant
radius throat geometry is to be approached in the large $\sigma$ limit and that
$f(\sigma)=0$ at extremality.

Using either metric, this amounts to the condition that
\eqn\form{R^2_T(d\sigma^2+d\Omega^2)} be the form of spatial metric, where
$R_T$ is the throat radius. This condition yields the first order differential
equation:
\eqn\conditions{(f^\prime)^2=(f^2-2Mf-l^2).}
This has a simple solution, (fixing an integration constant by requiring that
$f(\sigma)\to0$ at extremality):
\eqn\choice{f(\sigma)=\sqrt{M^2+l^2}\cosh\sigma+M.}

Recall that the  limiting procedure is to  take  $M,l\to0$ and $x\to\infty$
while holding constant $m=xM$ and $\lambda=xl$. To calculate the correctly
 scaling physics in this limit, set   $m=xM$ and $\lambda=xl$ together with
$M=\delta$ and $l=\delta{\lambda\over m}$ in all quantities. The extremal limit
is now simply $\delta\to0$. Quantities which survive  the extremal limit will
be those which scale with the correct powers of $\delta$.

The difference between the two dual cases arises when the throat radius is
calculated in the limit. In the case of \newmetrics\ it collapses:
\eqn\radii{R_T^2=(f^2+l^2)\to0}
while in the case of \newsigma\ it is non--zero:
\eqn\radiii{\hat{R}_T^2=(a^2+f_2^2)(f^2+l^2)
=f^2+2(x-1)Mf+(x-1)^2M^2+l^2x^2\to m^2+\lambda^2.}

Similarly, the rest of the metric \newmetrics\ vanishes in the limit while
 for the dual metric \newsigma:
\eqn\gtt{G_{\xi\xi}=-{(m^2+\lambda^2)\over
m^2}{\cosh^2\sigma-1\over(\cosh\sigma+\Delta)^2},} where
$\Delta=\sqrt{m^2+\lambda^2\over m^2}$

So the case of interest is clearly the dual solution of section 3, where the
string metric
remains finite and non--zero. After calculating the extremal
limit of the rest of the background fields by similar methods, the final
solution  for the extremal limit of the dual solution of section 3 in scaled
variables is (after rescaling $t$  by $1/m$):
\eqn\finalform{\eqalign{
d{\hat s}^2_S&=(m^2+\lambda^2)\left(d\sigma^2-{\cosh^2\sigma-1\over
(\cosh\sigma+\Delta)^2}(dt^2+2{\lambda\over m}\cos\theta
d\phi)^2+d\theta^2+\sin^2\!\theta\, d\phi^2\right);\cr
\e{-{\hat\Phi}}&=\cosh\sigma+\Delta;\cr
{\hat a}&=-{\lambda\over m}\cosh\sigma;\cr
\hat{A}_\phi&=2m\Delta\cos\theta {\Delta\cosh\sigma+1\over
\cosh\sigma+\Delta}\cr
\hat{A}_t&={2\lambda\over m}{\cosh\sigma\over\cosh\sigma+\Delta}\cr
\hat{B}_{\phi t}&={\lambda\Delta\cos\theta\over\cosh\sigma+\Delta}
}}

Notice that there was a shift in the dilaton and (exponentiated) axion by an
infinite constant in order to keep them finite in the limit. That this is an
allowed shift is easily seen by
looking at the low energy effective action written
in terms of the string metric $G_{\mu\nu}$ and
the scalar axion $a=e^{\rho}$. This is
\eqn\action{\eqalign{
I=\int d^4x\,\sqrt{-G}e^{-\Phi}\,&\left( R(G)+(\nabla\Phi)^2
-{1\over8}F^2\right.\cr
&\qquad\left.\ \ -{1\over2}e^{2(\Phi+\rho)}(\nabla\rho)^2
-{1\over16}e^{\Phi+\rho}
\epsilon^{\mu\nu\sigma\kappa}F_{\mu\nu}F_{\sigma\kappa}\right).\cr}}
This action (and the resulting equations of motion)
will be invariant under combined constant shifts
\eqn\shifti{
\Phi\rightarrow\Phi+c \qquad \rho\rightarrow\rho-c
}
or
\eqn\shiftii{
\eqalign{
e^{-\Phi}&\rightarrow (e^{-c})e^{-\Phi}=\hat{c}e^{-\Phi}\cr
a&\rightarrow (e^{-c})a=\hat{c}a\cr}
}
so the same constant can always be absorbed away into  the axion and dilaton.
This is indeed what was done to get the finite forms above.

This  extremal solution
\finalform\ is the same form as presented in ref.\cvj\ as  the low energy
limit of an exact conformal
field theory derived as a `heterotic coset' model. This is a consistent
combination of a $(0,2)$ supersymmetric
 $SL(2,\rline)\times SU(2)$
Wess--Zumino--Witten (WZW) model with a $U(1)\times U(1)$ subgroup gauged,
together with  a heterotic arrangement of
fermions\foot{In ref.\cvj, the gauge group was chosen as $U(1)\times U(1)$
where each factor contained identical theories. This is not an essential
choice, and was merely
the most symmetrical arrangement for the purposes of that paper.
It is consistent to truncate to just one copy of the  $U(1)$'s.}.
To match forms  exactly it is necessary to make a trivial
$U(1)$ gauge transformation.
Matching physical quantities defined there we get
\eqn\match{Q_A={m\hat{\lambda}\Delta\over2}\qquad
Q_B={m\Delta^2\over2}} which satisfy:
\eqn\satisfy{{Q_A^2\over\Delta^2-1}={Q_B^2\over\hat{\lambda}^2+1}
={Q_AQ_B\over\Delta\hat{\lambda}}}
where $\hat{\lambda}=\lambda/m$ and $\Delta=\sqrt{1+\hat{\lambda}^2}$ replace
the $\lambda$ and $\delta$, respectively,  of ref.\cvj. To compare dilatons it
 is necessary to rescale $\hat{\Phi}\to2\hat{\Phi}$.
Equations \satisfy\ are the precisely the relations  which arise from the
anomaly cancelation
conditions of ref.\cvj, ensuring that the model was conformally invariant.

Notice that the ${\hat\lambda}=0$ limit
corresponds to the extremal limit of the  magnetically charged dilaton black
hole of ref.\charged, which as a background of heterotic string theory was
shown to correspond to the low energy limit of
an exact conformal field theory in the first of
refs.\throat. This conformal field theory is a tensor product  of an orbifold
of an $SU(2)$ WZW and the solution of ref.\witten. This orbifold theory was
later shown in ref.\cvj\ to be equivalent to an example of the type of
`heterotic coset' introduced there, and is indeed a special case  of the
heterotic coset which gives rise to the extremal `Taub--NUT$+$throat'  solution
\finalform.

\newsec{Discussion.}

In this paper, we have presented new nontrivial solutions of the
low energy effective equations of motion in the heterotic string
theory. These solutions correspond to dyonic Taub--NUT spaces.
We have also confirmed the conjecture made in ref.\cvj\ that
the extremal limit of these Taub--NUT dyons corresponds
to the conformal field theory constructed there. This agreement
is not only in the string metric geometry, but also in the
dilaton, axion and gauge fields, all at one--loop in $\alpha^\prime$.
The exact conformal field theory
 provides all of the higher order (in $\alpha^\prime$) corrections
and also non--perturbative data. It is also satisfying to
find that the solutions naturally obey the constraints imposed
on the sigma model fields to ensure the vanishing of anomalies
in the gauged symmetries. The construction in ref.\cvj\ actually
introduces two background gauge fields, and we have set the
charges associated with one of these fields to zero in order
to make the comparison with the present solutions. A second
background gauge field
could be added by introducing a second ``chiral'' coordinate
in section 2. Then in addition to the boost in the gauge-time
directions made there, a $O(2)$ rotation amongst the ``chiral''
coordinates would introduce the desired second nontrivial gauge
field.

Both of the families of solutions given in sect.'s 2 and 3 are
dyonic carrying both magnetic and electric charges (although
we remind the reader that these charges are not defined in
a conventional way due to the topology of the solution). The
feature which truly distinguishes the two types of solutions
is the behavior of the scalar fields. Recall in sect.~2, the
dilaton charge was determined to be $D=-(x-1)M$ for the solutions
there. From eq.~\newaxidil, one finds a dilaton charge of
$\hat{D}=+(x-1)M$ for the solutions discussed in sect.~3.
Hence the string coupling $e^\Phi$ decreases in the central
region of the solutions in sect.~2, while it increases for
those in sect.~3.
One may think of the throat of the latter as being region of
strong coupling, while it is a region of weak coupling in
the solutions of sect.~2. These different behaviors are of
course expected since in general the $S$ duality transformation
which relates the two families transforms between strong and
weak coupling. Note that the axion charge also changes sign
under this transformation.

It would be a trivial matter to extend the family of
solutions constructed here by combining the $S$ and $T$
transformations in sect.~3. Equivalently, an integration
constant $c$ could be retained in eq.~\axion. Eq.'s
\newaxidil, \newsigma\ and \useful\ would then remain
unchanged although their explicit form would be slightly
more complicated. A further $SL(2,\rline)$ transformation
would be required to rescale the dilaton to restore $e^{-\hat{\Phi}}\to
1$ as $r\to\infty$. Then one finds that the resulting
electric and magnetic charge become
\eqn\Newcharges{\eqalign{
Q_E&=2\sqrt{x^2-1}{l+cM\over\sqrt{1+c^2}}\cr
Q_M&=-2\sqrt{x^2-1}{M-cl\over\sqrt{1+c^2}}\ .\cr}}
Hence the resulting solutions smoothly interpolate
between the two families presented here as $c$ varies
from 0 to $\infty$. Although either $Q_E$ or $Q_M$ can be set to
zero with an appropriate choice of $c$, the solution will
still contain both nontrivial electric and magnetic fields.

In the discussion so far, we have assumed that the boost parameter
$x\ge1$, but $x\le-1$ is also a valid range for this parameter.
In particular, the choice $x=-1$ corresponds to the famous discrete
duality transformation, which takes $G_{tt}\to {1/ G_{tt}}$.
For this particular choice, $G_{t\phi}$ vanishes. Thus the nontrivial
fibration of the time coordinate over the spatial two-sphere is
lost, and the topology of surfaces of constant $r$ is simply
$S^1\times S^2$. This change of topology under
a duality transformation was noted in a similar setting
by ref.\ref\topo{E. Alvarez, L. Alvarez-Gaume, J.L.F. Barbon and Y. Lozano,
{\sl Nucl.Phys.} {\bf B415} (1994) 71.}.

The geometry described by the original Taub--NUT metric
\taubnut\ is free of any curvature singularities (when $l\ne0$).
The present dyonic solutions will contain curvature singularities
at $\hat{r}_\pm=-{x-1\over2}M\pm\left[{(x-1)^2\over4}M^2-xl^2\right]^{1/2}$
where $f_2=0$. Thus no such singularities actually arise when
$l^2>{(x-1)^2\over4x}M^2$ (in which case $\hat{r}_\pm$ are complex).
In this aspect, the new solutions are similar to the
charged Taub--NUT solution of the Einstein-Maxwell
equations\brill, for which singularities arise when a
critical value of the charge is exceeded. In the present
solutions when $x>1$ (and $M>0$), the singularities will
occur at negative values of the radial coordinate, and so
are ``hidden" by the horizon-like surface at $r_+=M+(M^2+l^2)^{1/2}$.
In contrast for the solutions with $x<-1$ (and $M>0$),
one finds that $\hat{r}_+\ge r_+$, and so in this case the singularities
are not hidden.

\bigskip
\bigskip

{\bf Acknowledgments.}

The authors would like to thank Curt Callan, David Garfinkle and Frank Wilczek
for helpful remarks. The authors would also like to thank David Kastor
and Fernando Quevedo for useful comments on an earlier version of this
paper. CVJ is supported by a PPARC (UK) postdoctoral
fellowship. RCM is supported by NSERC of Canada, and Fonds FCAR du
Qu\'ebec.

\bigskip
\bigskip

\centerline{\bf Note Added}

After completion of the work described in this paper,
ref.\ref\kallosh{R. Kallosh, D. Kastor, T. Ortin and T. Torma, %
``Supersymmetry and Stationary Solutions in Dilaton-Axion Gravity,'' %
preprint hep-th/9406059.}
appeared, in which the solution of sections 2 and 3 is displayed.
That paper also notes that the extremal limit yields the `Taub--NUT$+$throat'
metric, corresponding to the exact conformal field theory of ref.\cvj.

\listrefs
\bye